\documentclass[aps,prb,twocolumn]{revtex4}

\usepackage{amsmath}
\usepackage{amssymb}
\usepackage{natbib}
\usepackage{graphicx}
\usepackage{bm}
\newcommand{\m}{\textbf{m}}

\def\be{\begin{equation}}
\def\ee{\end{equation}}
\def\ba{\begin{eqnarray}}
\def\ea{\end{eqnarray}}
\def\rr{{\bf r}}

\def\w{\omega}

\def\gyr{\mathrm{g}}
\def\e{\epsilon}
\def\ve{\varepsilon}
\def\D{\Delta}
\def\p{\partial}
\def\jj{\textbf{j}}
\def\pp{\textbf{p}}
\def\rr{\textbf{r}}
\def\qq{\textbf{q}}
\def\vv{\textbf{v}}

\def\EE{\textbf{E}}
\def\BB{\textbf{B}}

\def\d{\delta}

\def\Tr{\textrm{Tr}}

\def\bra{\langle}
\def\ket{\rangle}

\begin{document}

%\title{Plasmonic signatures in chiral magnetic effect}
\title{Kinetic orbital moments and nonlocal transport in disordered metals with nontrivial geometry}
%\title{Semiclassical theory of nonlocal transport in disordered metals with nontrivial geometry}
\date{\today}
\author{J. Rou, C. \c{S}ahin, J. Ma, D. A. Pesin}
\affiliation{Department of Physics and Astronomy, University of Utah, Salt Lake City, UT 84112 USA}
\begin{abstract}
We study the effects of spatial dispersion in disordered noncentrosymmetric metals. These include the kinetic magnetoelectric effect, natural optical activity of metals, as well as the so-called  dynamic chiral magnetic effect as a particular case of the latter. These effects are determined by the linear in the wave vector of an electromagnetic perturbation contribution to the conductivity tensor of a material, and stem from the magnetic moments of quasiparticles near the Fermi surface. We identify new disorder-induced contributions to these magnetic moments that come from the skew scattering and side jump processes, familiar from the theory of anomalous Hall effect.   We show that at low frequencies the spatial dispersion of the conductivity tensor comes mainly either from the skew scattering or intrinsic contribution, and there is always a region of frequencies in which the intrinsic mechanism dominates. Our results imply that in clean three-dimensional metals, current-induced magnetization is in general determined by impurity skew scattering, rather than intrinsic contributions. Intrinsic effects are expected to dominate in cubic enantiomorphic crystals with point groups $T$ and $O$, and in polycrystalline samples, regardless of their mobility.
\end{abstract}
%\pacs{}
\maketitle

\section{Introduction}

When a crystal is subjected to an \textit{ac} electromagnetic perturbation of frequency $\w$ and wave vector $\qq$, its linear response to the perturbation is fully described by the optical conductivity tensor, $\sigma_{ab}(\w,\qq)$.

Allowing for a moment for the possibility of time-reversal symmetry breaking in a crystal, denoted as the existence of magnetization $\bm M$, the optical conductivity tensor $\sigma_{ab}(\w,\qq;\bm M)$ can be written for small $M$ and $q$ as
\begin{equation}\label{eq:conductivity_general}
  \sigma_{ab}(\w,\qq;\bm M)\approx \sigma_{ab}(\w)+\chi_{abc}(\w)M_c+\lambda_{abc}(\w)q_c.
\end{equation}
In this expression, $\sigma_{ab}(\w)$ is the usual local optical conductivity, while the pseudotensor $\chi_{abc}$ and tensor $\lambda_{abc}$ describe optical activity of the crystal~\cite{LL8}, either due to the anomalous Hall effect ($\chi_{abc}$), or natural optical activity ($\lambda_{abc}$).
These tensors determine the antisymmetric part of the conductivity tensor. Indeed, the form of response is restricted by the Onsager relations~\cite{Melrose}:
\begin{equation}
  \sigma_{ab}(\w,\qq;\bm M)= \sigma_{ba}(\w,-\qq;-\bm M),
\end{equation}
which stem from the microscopic reversibility of the laws of physics, and imply that $\chi_{abc}$ and $\lambda_{abc}$ are antisymmetric with respect to the first pair of indices:
\begin{align}
  \begin{bmatrix}
  \lambda_{abc}\\
  \chi_{abc}
  \end{bmatrix}=-
  \begin{bmatrix}
  \lambda_{bac}\\
  \chi_{bac}
  \end{bmatrix}.
\end{align}

In this paper, we focus on the theory of the tensor $\lambda_{abc}(\w)$ in metals, at frequencies that are small compared to all band splittings at the Fermi surface. In other words, we consider the effects of spatial dispersion in disordered time-reversal invariant noncentrosymmetric metals.  We show that the linear-in-$\qq$ part of the conductivity tensor of these systems is of geometric origin. The term ``geometric origin'' is understood in the same way as the origin of the anomalous Hall effect, which is rooted in Berry phase physics~\cite{NagaosaReview}. As explained below, the results of this work represent the full theory of the natural optical activity in metals, and contain the theory of the dynamic chiral magnetic effect as a particular case.

The theory of the anomalous Hall effect (AHE), and thus the tensor $\chi_{abc}$, is now quite mature~\cite{NagaosaReview}. Importantly, since the seminal work of Haldane~\cite{Haldane2004}, it has been understood that (nonquantized part of) the low-frequency limit of this tensor is a Fermi surface property, originating from the Berry curvature in the band structure.  Berry phases at the Fermi surface not only determine the so-called intrinsic contribution to the anomalous Hall effect (AHE), but also can be shown to be behind the extrinsic -- side jump and skew scattering -- mechanisms~\cite{Sinitsyn2006}. By now the corresponding ideas have been well established, and excellent reviews have been written on the subject~\cite{SinitsynReview,NagaosaReview,XiaoNiu}

We show that mechanisms analogous to those responsible for the anomalous Hall effect determine the leading nonlocal correction to the conductivity tensor in time-reversal invariant disordered noncentrosymmetric metals. Specifically, we build the full theory of the tensor $\lambda_{abc}$ that determines the linear in wave vector part of the conductivity tensor~\eqref{eq:conductivity_general} in disordered metals. As a result, we identify two disorder-induced corrections to the orbital magnetic moment of quasiparticles that determine the extrinsic contribution to $\lambda_{abc}$.

From a physical point of view, the study is motivated by a recent revival in the interest in the geometric properties of the Fermi surface coming from two seemingly disconnected directions. One is the magnetohydrodynamics of the quark-gluon plasma\cite{kharzeevZhitnitsky2007,schafer2009, KharzeevSon2011,SonYamamoto,Gorbar2016} created in heavy-ion collisions in High Energy Physics, the other is the recent advent of Weyl semimetals in Condensed Matter Physics~\cite{Herring,WanVishwanath2011,BurkovBalents2011,Turner2013,hosur2013}. Both areas are related by the concept of the chirality, be it the intrinsic chirality of fermions in Lorentz-invariant quark-gluon plasma, or the chirality of conduction band electrons imparted by the chirality of the crystalline structure.

A particular example of a phenomenon connecting the two fields is the so-called chiral magnetic effect\cite{Vilenkin,Cheianov1998,Kharzeev2008,SonYamamoto}. One variety of it -- the dynamic Chiral Magnetic Effect (dCME)\cite{ChenBurkov2013,ChangYang2015,ChangNoWeyl,MaPesin2015,Zhong2016,Alavirad2015,Kharzeev2016} -- is defined as the electric current response to a slowly oscillating magnetic field, $\jj_{\textrm{cme}}(\w)=\gamma(\w)\BB(\w)$, where the chiral magnetic conductivity $\gamma(\w)$ is the response coefficient. One can easily see that the dCME is a particular case of Eq.~\eqref{eq:conductivity_general}. Indeed, it can be shown that in an isotropic clean metal $\lambda_{abc}\propto\epsilon_{abc}/\w$ at low frequencies, and Faraday's law $\qq\times \EE=\w \BB$ ensures that the current response described by such $\lambda_{abc}$ is exactly the dCME, establishing the relation.

The rest of the paper is organized as follows: Section~\ref{sec:results} contains the main results of the paper. Section~\ref{sec:kineq} lays out the kinetic equation formalism for the problem. In Section~\ref{sec:calculation} we apply the formalism of Section~\ref{sec:kineq} to calculate the nonlocal correction to the conductivity tensor in metals. Section~\ref{sec:physics} contains the description of the physical applications of the developed theory: natural optical activity and current-induced magnetization phenomena in disordered three-dimensional metals. Finally, in Section~\ref{sec:conclusions} we summarize our findings.

\section{Main results and qualitative considerations}\label{sec:results}
In this Section we summarize main results obtained in the rest the paper, and provide estimates for magnitudes of various quantities in a typical Tellurium-like helical metal. Most of the equations here will be repeated throughout the paper, often with more in-depth discussion.

The theory of the intrinsic contribution to the tensor $\lambda_{abc}$ in Eq.~\eqref{eq:conductivity_general} in crystals was developed in Refs.~\onlinecite{ChangYang2015,ChangNoWeyl,MaPesin2015,Zhong2016,Alavirad2015}, often under the disguise of considering the chiral magnetic effect at finite frequencies. It was realized that the intrinsic contribution was related to the so-called intrinsic orbital magnetic moment of quasiparticles, $\m^{\textrm{int}}_\pp$:\cite{noteonspin}
\begin{equation}\label{eq:magneticmoment}
\m^{\textrm{int}}_{\pp}=\frac {i\hbar e}2 \bra \p_\pp u_{\pp}|\times (h_\pp-\e_{\pp})|\p_\pp u_{\pp}\ket,
\end{equation}
where $|u_{\pp}\ket$ is the periodic part of the Bloch wave function (band index suppressed, only quasimomentum $\pp$ shown).
 This magnetic moment determined  $\lambda_{abc}$ via the gyrotropic tensor $g_{ab}$ dual to the latter, $\lambda_{abc}=\e_{abd}g_{dc}$, where
\begin{align}\label{eq:intg_mainresults}
    g_{ab}&=\frac{e}{(\w+\frac{i}\tau)} \int(d\pp)
  (m^{\textrm{int}}_{\pp a} \p_bf^0_\pp -\d_{ab}\m^{\textrm{int}}_{\pp}\cdot\p_\pp f^0_\pp).
\end{align}
Here $\tau$ is the transport scattering time, $f^0_\pp$ is the equilibrium distribution function of the electrons, $\p_a$ is the derivative with respect to the $a\textrm{th}$ component of the quasimomentum, and $(d\pp)\equiv d^3p/(2\pi \hbar)^3$.

In this paper, we show that in the presence of disorder, the tensor $\lambda_{abc}$ still stems from an appropriately defined magnetic moment of quasiparticles, which appears only under nonequilibrium conditions. There are two contributions to such \emph{kinetic} magnetic moment:  skew scattering\cite{smit1958,mott1965} from impurities, and the coordinate shift upon impurity scattering (the ``side jump'' effect\cite{berger1970,nozieres1973,Belinicher1982,Sinitsyn2006}). First, we present the final expression for these magnetic moment contributions, and then discuss the quantities that enter into them. The skew scattering contribution is given by
\begin{equation}\label{eq:skmomentresults}
  \m^{\textrm{sk}}_\pp=\frac{e\tau^2}{(1-i\w\tau)^2}\bm Q^{\textrm{sk}}_\pp\times\p_{\pp}\e_{\pp},
\end{equation}
and the side jump one is
\begin{equation}\label{eq:sjmomentresults}
  \m^{\textrm{sj}}_\pp=\frac{e\tau}{1-i\w\tau}\vv^{\textrm{sja}}_{\pp} \times\p_\pp\e_\pp.
\end{equation}
In these expressions $\p_\pp\e_\pp$ is the usual group velocity of electrons in a given band, $\bm Q^{\textrm{sk}}_\pp$ we will refer to as skew acceleration, and $\vv^{\textrm{sja}}_\pp$ is the side jump accumulation velocity familiar from the anomalous Hall effect context (see Ref.~\onlinecite{SinitsynReview} for a review). The expressions for $\bm Q^{\textrm{sk}}_\pp$ and  $\vv^{\textrm{sja}}_\pp$ involve the symmetric and antisymmetric parts of the impurity scattering probability from $\pp'$ to $\pp$, $w^{\textrm{S,A}}_{\pp\pp'}$, see Eqs.~\eqref{eq:transitionprobability} and~\eqref{eq:symantisym}, as well as the coordinate shift of the electron in the same transition, $\d\rr_{\pp\pp'}$, Eq.~\eqref{eq:sidejump}. Specifically,

\begin{equation}
  \bm Q^{\textrm{sk}}_\pp=\int (d\pp')w^{\textrm{A}}_{\pp\pp'}
\d(\e_{\pp}-\e_{\pp'})(\p_{\pp}\e_{\pp}-\p_{\pp'}\e_{\pp'}),
\end{equation}
and
\begin{equation}
  \vv^{\textrm{sja}}_\pp=\int (d\pp')w^{\textrm{S}}_{\pp\pp'}\d \rr_{\pp\pp'} \delta\left(\e_\pp-\e_{\pp'}\right).
\end{equation}
The physical meaning of $\bm Q^{\textrm{sk}}_\pp$ and  $\vv^{\textrm{sja}}_\pp$ is clear from their definitions: $\vv^{\textrm{sja}}_\pp$ is a disorder-induced velocity of a wave packet due to the accumulation of the side jump events; in turn, $\bm Q^{\textrm{sk}}_\pp$ is a disorder-induced acceleration of a wave packet due to the velocity change accumulation upon skew scattering.

These considerations make apparent the origin of extrinsic contributions to the magnetic moments, Eqs.~\eqref{eq:skmomentresults} and~\eqref{eq:sjmomentresults}: the product of $\bm Q^\textrm{sk}_\pp$ and $\min[1/\w,\tau]^2$, as well as the product $\vv^{\textrm{sja}}_{\pp} \min[1/\w,\tau]$ have the meaning of an average displacement of a charge carrier due to accumulation of either skew scattering, or side jump events. The cross products of these quantities with the group velocity $\p_\pp\e_\pp$ of the carrier determine the contribution of the extrinsic effects to the effective ``kinetic'' magnetic moment of a quasiparticle.

Finally, introducing the total magnetic moment $\m^{\textrm{tot}}_{\pp}=\m^{\textrm{int}}_{\pp}+\m^{\textrm{sk}}_{\pp}+\m^{\textrm{sj}}_{\pp}$, the gyrotropic tensor is written as
\begin{equation}\label{eq:gyrototalresults}
  g_{ab}= \frac{e}{(\w+\frac{i}{\tau})}\int(d\pp)
  (m^{\textrm{tot}}_{\pp a} \p_bf^0_\pp -\d_{ab}\m^{\textrm{tot}}_{\pp}\cdot\p_\pp f^0_\pp).
\end{equation}
Equations Eqs.~\eqref{eq:skmomentresults}, \eqref{eq:sjmomentresults}, and~\eqref{eq:gyrototalresults} are the main results of this work. Using these findings, we will describe the phenomena of natural optical activity and kinetic magnetoelectric effect in metals in Sections~\ref{sec:NOA} and~\ref{sec:kineticME}.
\subsection{Estimates for a typical helical  metal}
\begin{figure}
  \centering
  \includegraphics[width=2in]{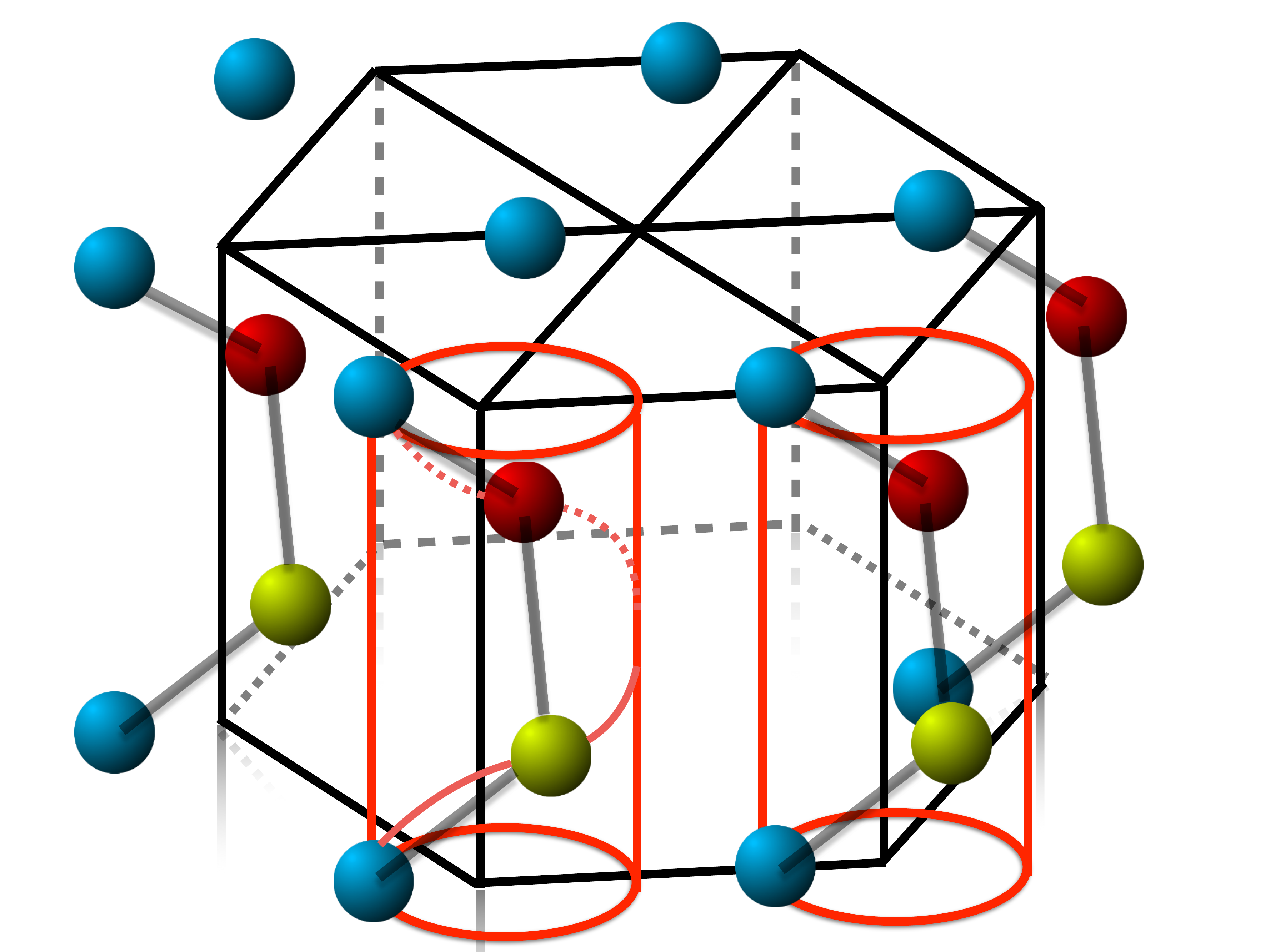}
  \caption{(Color online) Tellurium-type crystal structure. Atomic helices wind around the vertical edges of the hecagonal prism representing the unit cell of the crystal.}\label{fig:Te}
\end{figure}

We conclude this Section with a brief qualitative discussion of the origin of the skew scattering and side jump effects in a typical helical metal. To have a clear physical picture, as well as an opportunity to apply the obtained results to a concrete system, we consider a specific crystal structure of the hexagonal Tellurium, Fig.~\ref{fig:Te}, which is one of the most important enantiomorphic crystals.

We first consider skew scattering. It is well known\cite{mott1965} that in the case of a free electron with spin $\bm s$ scattering off a spin-orbit coupled impurity, the scattering probability contains an antisymmetric contribution $\w^\textrm{A}_{\pp\pp'}\propto \bm s\cdot\pp\times\pp'$. For an electron in a noncentrosymmetric crystal, in which the band structure is spin-split, one faces the question of what plays the role of the spin $\bm s$. Fig.~\ref{fig:Te} helps answer it: it is clear that an electron propagating along the c axis -- that is, along the helices -- obtains a magnetic moment along the helix, which is proportional to the momentum along the helix, at least for small momenta, $\m_\pp\propto p_z \bm z$. It is then plausible to assume that $\m_\pp$ plays the role of $\bm s$ in this situation, and the properly symmetrized antisymmetric probability has the form of $\w^\textrm{A}_{\pp\pp'}\propto (p_z+p'_z)\bm z\cdot\pp\times\pp'$. This form satisfies the time-reversal condition $\w^\textrm{A}_{\pp,\pp'}=\w^\textrm{A}_{-\pp',-\pp}$, and is in fact dictated by the symmetry of the lattice. It can also be confirmed with a microscopic calculation\cite{RouUnpublished}.

 To calculate the skew acceleration, we assume an isotropic spectrum with a density of states $\nu_\pp$, and introduce the skew scattering time $\tau_{\textrm{sk}}$ via
\begin{align}
  \w^\textrm{A}_{\pp\pp'}=\frac{1}{\nu_\pp\tau_{\textrm{sk}}}
  (\bm z\cdot \bm e_\pp+\bm z\cdot \bm e_{\pp'})\bm z\cdot \bm e_\pp\times\bm e_{\pp'},
\end{align}
where $\bm e_\pp$ is the unit vector along $\pp$. Then we obtain for the skew acceleration
\begin{equation}
  \bm Q^{\textrm{sk}}_\pp=-\frac{\p_p\e_p}{3\tau_{\textrm{sk}}} (\bm z\cdot\bm e_\pp)\bm z\times\bm e_\pp,
\end{equation}
and the skew scattering contribution to the magnetic moment
\begin{equation}\label{eq:skewestimate}
  \m^{\textrm{sk}}_\pp=\frac{1}{3}\frac{e\tau^2}{(1-i\w\tau)^2}\frac{(\p_p\e_p)^2}{\tau_{\textrm{sk}}} (\bm z\cdot\bm e_\pp)\bm e_{\pp}\times\bm z\times\bm e_\pp.
\end{equation}
We will calculate the corresponding contribution to the gyrotropic tensor at zero temperature, and will use $v_F$ and $\nu_F$ denote the Fermi velocity and the density of states at the Fermi level, respectively. Then
\begin{align}\label{eq:skgyroresults}
  g^{\textrm{sk}}_{ab}=\frac{1}{45}\frac{e^2\tau^3}{i(1-i\w\tau)^3}\frac{\nu_F v^3_F}{\tau_{\textrm{sk}}}
  \begin{pmatrix}
    1&0&0\\
    0&1&0\\
    0&0&-2
  \end{pmatrix}.
\end{align}
Note that the tracelessness of the gyrotropic tensor is a general property of the extrinsic contributions described above (see the end of Section~\ref{sec:symmetry} for further details).

As far as the side jump mechanism is concerned, there exist extensive discussions of the physics behind the side jump effect itself\cite{nozieres1973,Sinitsyn2006,engel2007}, which we will not repeat here. To estimate the side jump contribution to the gyrotropic tensor, we consider a weak spherically-symmetric impurity potential, which is smooth on the scale of the lattice spacing, producing mostly small-angle scattering. In this case the side jump length is independent of the specific form of the potential, and is fully determined by the Berry curvature of the band, $\bm \Omega_\pp$:\cite{Sinitsyn2006}
\begin{align}
  \bm \Omega_\pp=i\hbar \bra\p_\pp u_\pp|\times|\p_\pp u_\pp\ket.
\end{align}
The coordinate shift $\d\rr_{\pp\pp'}$ is given by
\begin{equation}
  \d\rr_{\pp\pp'}=\bm \Omega_{\frac{\pp+\pp'}2}\times (\pp-\pp').
\end{equation}
We take $\bm \Omega_\pp\propto p_z \bm z$ as appropriate for the upper valence band of Tellurium to obtain
\begin{align}
  \d\rr_{\pp\pp'}=a_{\textrm{sj}}(\bm z\cdot \bm e_\pp+\bm z\cdot \bm e_{\pp'})\bm z\times (\bm e_\pp-\bm e_{\pp'}).
\end{align}
In the above expression $a_{\textrm{sj}}$ sets the overall scale of the side jump magnitude. Given the qualitative nature of present considerations, we calculate the side jump accumulation velocity setting $w^{\textrm{S}}_{\pp\pp'}=1/\nu_\pp\tau$:
\begin{equation}
  \vv^{\textrm{sja}}_\pp=\frac{a_{\textrm{sj}}}{\tau}(\bm z\cdot \bm e_\pp)\bm z\times\bm e_\pp.
\end{equation}
The expression for the side jump magnetic moment is
\begin{equation}\label{eq:sjmomenttellurium}
  \m^{\textrm{sj}}_\pp=-\frac{e a_{\textrm{sj}}\p_p\e_p}{1-i\w\tau}(\bm z\cdot \bm e_\pp)\bm e_\pp\times \bm z\times\bm e_\pp,
\end{equation}
and the contribution to the gyrotropic tensor is
\begin{align}\label{eq:sjgyroresults}
  g^{\textrm{sj}}_{ab}=-\frac{1}{15}\frac{e^2\tau}{i(1-i\w\tau)^2}a_{\textrm{sj}}\nu_F v^2_F
  \begin{pmatrix}
    1&0&0\\
    0&1&0\\
    0&0&-2
  \end{pmatrix}.
\end{align}

Finally, the intrinsic contribution comes from the intrinsic magnetic moment, which for small $p_z$ can be written as \cite{Farbshtein2012}
\begin{equation}\label{eq:intmomentresults}
  \m^{\textrm{int}}_\pp=\mu_{\textrm{int}}\bm z(\bm z\cdot \bm e_\pp).
\end{equation}
The intrinsic part of the gyrotropic tensor is thus
\begin{align}\label{eq:intgyroresults}
  g^{\textrm{int}}_{ab}=-\frac{1}{3}\frac{e\mu_{\textrm{int}}\tau}{i(1-i\w\tau)}\nu_Fv_F
  \begin{pmatrix}
    1&0&0\\
    0&1&0\\
    0&0&0
  \end{pmatrix}.
\end{align}
Eqs.~\eqref{eq:skgyroresults},~\eqref{eq:sjgyroresults}, and~\eqref{eq:intgyroresults} allow to draw some general conclusions about the contributions of extrinsic effects to various physical effects. For instance, the polarization rotation and circular dichroism for light propagating along the optic axis of the crystal (cf. Section~\ref{sec:NOA}) are determined by $g_{zz}$. It is clear from Eq.~\eqref{eq:intgyroresults} that at frequencies small compared to relevant band splittings, where the current theory is applicable,  these effects are dominated by the extrinsic effects, regardless of the mobility of a sample. This should be contrasted with the case of optical frequencies\cite{IvchenkoTe}, in which $g_{zz}$ is fully determined by the band structure. On the other hand, both extrinsic and intrinsic effects contribute to the current-induced magnetization (Section~\ref{sec:kineticME}).

It is also interesting to compare the typical values of matrix elements of the gyrotropic tensor that come from various mechanisms.  For $\w\tau\ll 1$, one has
\begin{subequations}\label{eq:estimatesresults}
  \begin{align}
    g^{\textrm{sk}}&\sim e^2\tau^3\frac{\nu_F v^3_F}{\tau_{\textrm{sk}}}\sim \frac{e^2}{\hbar}k^2_F\ell^2\frac{\tau}{\tau_{\textrm{sk}}},\label{eq:skloww}\\
    g^{\textrm{sj}}&\sim e^2\tau a_{\textrm{sj}}\nu_F v^2_F\sim\frac{e^2}{\hbar}k_F^2\ell a_{sj},\\
    g^{\textrm{int}}&\sim e\mu_{\textrm{int}}\tau\nu_Fv_F\sim \frac{e^2}{\hbar}k_F^2\ell \frac{\mu_{\textrm{int}}}{e v_F}.\label{eq:intloww}
  \end{align}
\end{subequations}
The first similarity sign (``$\sim$'') pertains to the considered simplified model of a helical metal; the expressions following the second one, expressed through the Fermi wave vector $k_F$, and the elastic mean free path $\ell$ hold more generally, and allow a comparison of extrinsic and intrinsic mechanism in various systems. Such comparison will be carried out in more detail in Sections~\ref{sec:NOA} and~\ref{sec:kineticME}. Here we only note that for clean systems with large enough value of $k_F\ell$, skew scattering always dominates the physical phenomena related to the nonlocality of the conductivity tensor.

What is interesting in the present case is the fact that the different frequency dependence of the extrinsic and intrinsic contribution allows the intrinsic ones to overpower skew scattering (the side jump contribution appears to be small in all realistic examples\cite{RouUnpublished}) at frequencies still within the applicability range of the theory. Consider, for instance, the case of a Weyl semimetal, in which $\mu_{\textrm{int}}\sim e v_F/k_F$, and the intrinsic effects are large. Comparing Eq.~\eqref{eq:skgyroresults} and Eq.~\eqref{eq:intgyroresults}, and using Eqs.~\eqref{eq:skloww} and~\eqref{eq:intloww}, one can easily see that for $\w\gg \w_0\equiv \sqrt{k_F\ell/\tau\tau_{\textrm{sk}}}$ the intrinsic effects dominate the skew scattering. The crossover frequency $\w_0$ satisfies $\hbar\w_0/\e_F\sim\sqrt{\hbar/\e_F\tau_{\textrm{sk}}}\ll1$, hence the crossover regime lies well within the applicability region of the present theory.

\section{Semiclassical transport theory}\label{sec:kineq}

In this Section, we lay out the framework to describe the response of a crystal to a weak electromagnetic perturbation varying slowly in space and time, which can be described with an effective single-band Boltzmann equation. In this equation, the effective velocity of quasiparticles, the force that acts on them, as well as the impurity collision integral are modified by the effects of interband coherence\cite{SinitsynReview}.  The goal of this Section is to describe this type of Boltzmann equation in the context of non-local conductivity calculation in metals.

For the clean metal case, a detailed account of the kinetic equation formalism for a non-local conductivity calculation was laid out in Ref.~\onlinecite{MaPesin2015}. Here we briefly repeat the relevant equations, and in addition,  introduce the appropriate form of the impurity collision integral to treat disorder.

The main results of this Section are Eqs.~\eqref{eq:collintW},~\eqref{eq:collintM}, and ~\eqref{eq:collintSJ} for contributions to the impurity collision integral in the Boltzmann equation, as well as Eqs.~\eqref{eq:qpcurrent},~\eqref{eq:mcurrent}, and ~\eqref{eq:sjcurrent} for the electric current in the presence of impurities.

The semiclassical kinetic equation for the distribution function $f_{n\pp}$ of band $n$ and momentum $\pp$ has the standard form:
\begin{equation}\label{eq:kineqgeneral}
  \p_t f_{n\pp}+\dot\rr\p_\rr f_{n\pp}+\dot\pp\p_\pp f_{n\pp}=I_{\textrm{st}},
\end{equation}
where $I_{\textrm{st}}$ is the collision integral.

The expressions for the velocity, $\dot \rr$, of a quasiparticle, and its quasimomentum rate of change, $\dot \pp$, are modified by the effects of interband coherence. These come from several sources, and are listed below.

Interband coherence induced by the acceleration due to the Lorentz force manifests itself in the appearance of the Berry curvature ${\bf \Omega}_{n\pp}$ of band $n$. It can be expressed using the periodic parts of the Bloch wave functions $|u_{n\pp}\ket$ of band $n$, and is given by
\begin{align}\label{eq:Omega}
{\bf \Omega}_{n\pp}=i\hbar\bra\p_\pp u_{n\pp}|\times|\p_\pp u_{n\pp}\ket.
\end{align}

Spatial gradients of the distribution function of band $n$ induce coherence that leads to the appearance of the orbital magnetic moment of quasiparticles \cite{noteonspin},
\begin{equation}\label{eq:magneticmoment}
\m_{n\pp}=\frac {i\hbar e}2 \bra \p_\pp u_{n\pp}|\times (h_\pp-\e_{n\pp})|\p_\pp u_{n\pp}\ket,
\end{equation}
where $h_{\pp}$ is the Bloch Hamiltonian, and $\epsilon_{n\pp}$ is the energy dispersion of band $n$.

The presence of this magnetic moment modifies the expression for the band energy dispersion,
\begin{equation}\label{eq:qpenergy}
  E_{n\pp}=\e_{n\pp}-\m_{n\pp}\BB,
\end{equation}
as well as the band velocity,
\begin{equation}\label{eq:qpvelocity}
  \vv_{n\pp}=\p_{\pp}E_{n\pp}.
\end{equation}

Using these quantities, the semiclassical equations of motion for band $n$ can be written as\cite{XiaoNiu}
\begin{eqnarray}\label{eq:EOM}
  \dot\rr&=&\vv_{n\pp}-\dot\pp\times{\bf \Omega}_{n\pp},\nonumber\\
  \dot\pp&=&e\EE+e\dot\rr\times\BB.
\end{eqnarray}
The corresponding standard expressions for the velocity and the momentum rate of change are
\begin{eqnarray}\label{eq:qpspeed}
  \dot\rr&=&\frac{1}{D_{\BB}}\left(\vv_{n\pp}-e\EE\times{\bf \Omega}_{n\pp}-e(\vv_{n\pp}\cdot{\bf \Omega}_{n\pp})\BB\right),\nonumber\\
  \dot\pp&=&\frac{1}{D_{\BB}}\left(e\EE+e\vv_{n\pp}\times\BB-e^2(\EE\cdot\BB){\bf \Omega}_{n\pp}\right),\nonumber\\
  D_\BB&=&1-e\BB{\bf \Omega}_{n\pp}.
\end{eqnarray}
The equation for $\dot\rr$ contains the group velocity $\vv_{n\pp}$, as well as the anomalous velocity due to the electric and magnetic parts of the Lorentz force~\cite{XiaoNiu}; in particular, the last term in brackets is commonly associated with the static CME~\cite{Vilenkin,Nielsen1983,Cheianov1998,Kharzeev2008,Son2013,Niu2013}, but plays no role in the present discussion.  The equation for $\dot\pp$, besides the usual Lorentz force, contains an ``$\EE\cdot\BB$'' term, which describes the chiral anomaly at the quasiclassical level.\cite{SonSpivak2013,WongTserkovnyak2011}

The factor $D_{\BB}$ is the phase space measure that appears due to variables $\rr$ and $\pp$ not being a canonical pair\cite{XiaoNiu2005,Bliokh2006,Duval2006}. It can be shown that the $1/D_{\BB}$ prefactor in the expressions for the velocities in real and momentum spaces is compensated by $D_{\BB}$ appearing in the momentum integrals over the Brillouin zone, and does not affect the linear response. Hence, we will omit it from now on.

\begin{widetext}
We now turn to the description of the disorder effects. We start with the discussion of the collision integral, while the question of velocity renormalization by the disorder is treated below, see Eq.~\eqref{eq:sjcurrent}.

The central quantity describing a collision of an election with an impurity is the T-matrix of the latter, $\hat T_{\textrm{imp}}$. Below, the matrix elements of the T-matrix are defined as
\begin{equation}
T_{n\pp,n'\pp'}=V\bra n\pp|\hat T|n'\pp'\ket,
\end{equation}
in which $V$ is the volume of the system, $|n\pp\ket=\frac{1}{\sqrt{N}}\exp(i\pp\rr/\hbar)|u_{n\pp}\ket$ are Bloch states normalized to unity, and indices must be read from right to left to infer the transition direction.

For a randomly distributed ensemble of impurities with impurity concentration $n_i$, the probability of the $n'\pp'\to n\pp$ transition is given by
\begin{equation}\label{eq:transitionprobability}
  W_{n\pp,n'\pp'}=\frac{2\pi}{\hbar}\frac{1}{V}n_i|T_{n\pp,n'\pp'}|^2\d(\e_{n\pp}-\e_{n'\pp'})\equiv \frac{1}{V}w_{n\pp,n'\pp'}\d(\e_{n\pp}-\e_{n'\pp'}).
\end{equation}
It is important for us that in a noncentrosymmetric crystal, the transition rate is not symmetric with respect to the interchange of the initial and final states, $W_{n\pp,n'\pp'}\neq W_{n'\pp',n\pp}$\cite{LL3}. Therefore, $w_{n\pp,n'\pp'}$ in Eq.~\eqref{eq:transitionprobability} can be written as a sum of symmetric and antisymmetric components:
\begin{equation}\label{eq:symantisym}
  w_{n\pp,n'\pp'}=w^{\textrm{S}}_{n\pp,n'\pp'}+w^{\textrm{A}}_{n\pp,n'\pp'},\quad
   w^{\textrm{S,A}}_{n\pp,n'\pp'}=\pm w^{\textrm{S,A}}_{n'\pp',n\pp},
\end{equation}
The antisymmetric part $w^A$ is responsible for the skew scattering.

The conservation of the probability flux in impurity scattering (the optical theorem) imposes a restriction on the asymmetric part of the scattering rate. Indeed, probability conservation implies that the total transition rate out of a given state is equal to the total transition rate into that state from all other states\cite{Sturman1984}. In other words, the following sum rule holds:
\begin{equation}
  \sum_{n\pp}W_{n\pp,n'\pp'}=\sum_{n\pp}W_{n'\pp',n\pp}.
\end{equation}
This immediately gives
\begin{equation}\label{eq:unitarity}
 \sum_{n}\int (d\pp) w^{\textrm{A}}_{n\pp,n'\pp'}\d(\e_{n\pp}-\e_{n'\pp'})=0,
\end{equation}
where $(d\pp)=d^3p/(2\pi \hbar)^3$.

Finally, we note that since the position of the center of a wave packet within the unit cell depends on its quasimomentum, the sudden change of the quasimonentum upon impurity scattering will lead to a shift in the wave packet location. This coordinate shift - commonly called ``side jump'' - is given for a weak impurity potential $V^{imp}$ by\cite{Belinicher1982,Sinitsyn2006}
\begin{equation}\label{eq:sidejump}
  \d\rr_{n\pp,n'\pp'}=i\hbar \bra u_{n\pp}|\p_\pp u_{n\pp}\ket-i\hbar \bra u_{n'\pp'}|\p_{\pp'} u_{n'\pp'}\ket
  -\hbar(\p_\pp+\p_{\pp'})\textrm{arg}(V^{imp}_{n\pp,n'\pp'}).
\end{equation}
In the presence of electromagnetic fields, one must take into account the work done by the electric field as an electron gets displaced within the unit cell, as well as the change in the Zeeman energy, Eq.~\eqref{eq:qpenergy}, during the collision. These considerations lead to modifications\cite{noteonunitarity} in the energy-conserving $\d$-function in Eq.~\eqref{eq:transitionprobability}, and the collision integral can be generally written as
\begin{equation}
  I_{\textrm{st}}=-\sum_{n'}\int (d\pp')D_\BB (w_{n'\pp',n\pp}f_{n\pp}-w_{n\pp,n'\pp'}f_{n'\pp'})\d(E_{n\pp}-E_{n'\pp'}-e\EE\d\rr_{n\pp,n'\pp'}),
\end{equation}
 The phase space measure $D_\BB$ will be omitted from now on, as it does not alter the linear response. We note in passing that the accumulation of the subsequent side jumps also modifies the wave packet velocity, and this effect is discussed further below.

To linear order in the electromagnetic fields, the impurity collision integral has four parts, $I_{\textrm{st}}=I^S+I^A+I^{\textrm{m}}+I^{\textrm{sj}}$, where
$I^{S,A}$ are symmetric and antisymmetric parts of the usual collision integral,
\begin{equation}\label{eq:collintW}
  I^{\textrm{S,A}}=-\sum_{n'}\int (d\pp')D_\BB (w^{\textrm{S,A}}_{n'\pp',n\pp}
  \d f_{n\pp}-w^{\textrm{S,A}}_{n\pp,n'\pp'}\d f_{n'\pp'})\d(\e_{n\pp}-\e_{n'\pp'}),
\end{equation}
 in which $\d f_{n\pp}\sim O(\EE,\BB)$;  $I^{\textrm{m}}$ is an effective generation term due to the Zeeman energy change during a collision:
 \begin{equation}\label{eq:collintM}
   I^{\textrm{m}}=-\p_{\e_{n\pp}} f^0_{n\pp}\sum_{n'}\int (d\pp') w^{\textrm{S}}_{\pp,\pp'}(\m^{\textrm{int}}_\pp-\m^{\textrm{int}}_{\pp'})\BB\delta\left(\e_{n\pp}-\e_{n'\pp'}\right).
 \end{equation}

Finally, the side jump contribution $I^{\textrm{sj}}$, which also plays the role of a generation term, is related to the work done by the electric field as an electron completes a side jump~\cite{SinitsynReview}:
\begin{eqnarray}\label{eq:collintSJ}
  I^{\textrm{sj}}&=&-\p_{\e_{n\pp}} f^0_{n\pp}\sum_{n'}\int (d\pp')w^{\textrm{S}}_{n\pp,n'\pp'}e\EE\d\rr_{n\pp,n'\pp'}\delta\left(\e_{n\pp}-\e_{n'\pp'}\right). \end{eqnarray}

For solenoidal fields, the standard interpretation suffers from the fact that the work done by the electric field depends on the integration path due to non-zero magnetic field. However, the difference of between the ``straight path'' expression, $e\EE\d \rr_{\pp\pp'}$, and any other path is roughly determined by the time derivative of the magnetic field flux through the scattering region, and is negligible.

To complete the kinetic scheme, we have to write down the equation for the electric current. In the present case, there are three contributions,
\begin{equation}\label{eq:totalcurrent}
  \jj=\jj_{\textrm{qp}}+\jj_{\textrm{m}}+\jj_{\textrm{sj}},
\end{equation}

The first one, $\jj_{\textrm{qp}}$, comes from the wave packet velocity of Eq.~(\ref{eq:qpspeed}), and is given by
\begin{equation}\label{eq:qpcurrent}
  \jj_{\textrm{qp}}=e\sum_{n}\int (d\pp)\left(\p_\pp\e_{n\pp}-\p_\pp(\m^{\textrm{int}}_{n\pp}\BB)-e\EE\times{\bf \Omega}_{n\pp}-e(\p_\pp\e_{n\pp}\cdot{\bf \Omega}_{n\pp})\BB\right) f_{n\pp}.
\end{equation}
The second term in Eq.~\eqref{eq:totalcurrent} is the magnetization current due to the existence of the intrinsic orbital moment of quasiparticles~\cite{WongTserkovnyak2011,PesinMacDonaldMonopole,Son2013,MishchenkoStarykh2014}:
\begin{equation}\label{eq:mcurrent}
  \jj_{\textrm{m}}=\nabla_\rr\times\sum_{n}\int (d\pp)\m^{\textrm{int}}_{n\pp} f_{n\pp}.
\end{equation}
Finally, the last contribution to the total current, $\jj_{\textrm{sj}}$, is the current due to the side jump accumulation, given by
\begin{equation}\label{eq:sjcurrent}
  \jj_{\textrm{sj}}(\w,\qq)=e\sum_{nn'}\int(d\pp)(d\pp')w^{\textrm{S}}_{n\pp,n'\pp'}\d\rr_{n\pp,n'\pp'} \delta\left(\e_{n'\pp'}-\e_{n\pp}\right)f_{n'\pp'}.
\end{equation}
Qualitatively, it comes from the net displacement accumulated as an electron completes a series of side jumps. Formally, it stems from the interband coherence induced by collisions with impurities.

The kinetic equation~\eqref{eq:kineqgeneral}, together with the collision integrals~\eqref{eq:collintW} and~\eqref{eq:collintSJ}, and the expression for the current~\eqref{eq:totalcurrent} constitute the full kinetic scheme for the problem at hand.

The intrinsic orbital moment of quasiparticles, $\m^{\textrm{int}}_{n\pp}$, the antisymmetric part of the impurity scattering probability determined by $w^{\textrm{A}}_{n\pp.n'\pp'}$, and the side jump displacement, $\d\rr_{n\pp,n'\pp'}$, lead to the intrinsic, skew scattering, and side jump mechanisms of nonlocal response, respectively. They all stem from the existence of Berry-Pancharatnam phases in the band structure\cite{Sinitsyn2006}, and hence are geometric in nature. Since these effects are relatively small, the corresponding contributions to $\lambda_{abc}$, Eq.~\eqref{eq:conductivity_general}, are additive and can be considered independently, which is done in the subsequent Sections. For the sake of brevity, we will suppress the band index $n$ from now on.
% It can be reinstated by trivially substituting $\pp\to n\pp$ in all indices.

\section{Nonlocal transport in metals}\label{sec:calculation}
In this Section, we use the formalism described in Section~\ref{sec:kineq} to calculate tensor $\lambda_{abc}$ that determines linear-in-$\qq$ effects of spatial dispersion in the conductivity tensor, Eq.~\eqref{eq:conductivity_general}. The main results of this Section are Eqs.~\eqref{eq:intrinsicfinal},~\eqref{eq:skewfinal}, and~\eqref{eq:sjfinal} for intrinsic, skew scattering, and side jump contributions to tensor $\lambda_{abc}$.

\subsection{Intrinsic contribution in a disordered metal}\label{sec:intrinsic}

We first consider the contribution to $\lambda_{abc}$ that comes solely from the $\pp$-dependence of the periodic parts of Bloch functions, via the intrinsic orbital moment of quasiparticles, $\m_{\pp}$. To this end, we neglect the skew scattering and side jump effects by setting $w^\textrm{A},\d\rr\to 0$, and call this contribution ``intrinsic'', even though it does depend on the momentum relaxation time in a band, or $w^S$.

Since we are interested in the linear response to an electromagnetic field, it is sufficient to consider monochromatic fields, $\EE,\BB\propto \exp\left(i\qq\rr-i\w t\right)$. The distribution function response is then also monochromatic, and the kinetic equation has the following form:
%\begin{equation}\label{eq:intrinsickineq}
%-i\w f_\pp+i\qq\p_\pp \e_\pp f_\pp+e\EE\p_\pp f^0_\pp=-\int (d\pp') W^{s}_{\pp,\pp'}(f_{\pp}-f_{\pp'})\d(\e_\pp-\m^{\textrm{int}}_\pp\BB-\e_{\pp'}+\m_{\pp'}\BB).
%\end{equation}
\begin{equation}\label{eq:intrinsickineq}
-i\w f_\pp+i\qq\p_\pp \e_\pp f_\pp+e\EE\p_\pp f^0_\pp=I^\textrm{S}+I^{\textrm{m}}.
\end{equation}
The expression for the current contribution which is linear-in-$\qq$ is
\begin{equation}
   \jj_g^{\textrm{int}}(\w,\qq)=e\int (d\pp)\m^{\textrm{int}}_\pp\BB\p_\pp f^0_\pp+e\int (d\pp)\p_\pp\e_\pp f^{\BB}_\pp+\int(d\pp)i\qq\times\m^{\textrm{int}}_\pp f^\EE_\pp.
\end{equation}
The subscript $g$ stands for ``gyrotropic''.  The first term on the right hand side is the current due to renormalization of the velocity expectation value for state $\pp$ in the presence of magnetic field; the second term is the usual ballistic current calculated using the nonequilibrium distribution function, $f^{\BB}_\pp$, due to coupling to a time-dependent magnetic field, and the last term is the nonequilibrium magnetization current, appearing due to nonuniform acceleration of electrons by the electric field, described by $f^{\EE}_\pp$. Because we work to linear order in the electric field gradients, the terms in $\jj_g^{\textrm{int}}$ containing $\qq$ can be dropped next to the magnetic field; the same is true for $f^\EE_\pp$, since the magnetization current contains $\qq$ in its definition.

To proceed, we separate the non-equilibrium part of the distribution function $\d f_\pp\equiv g_\pp\p_{\e_\pp} f^0_\pp$, and define an integral operator $\hat {\cal L}$ that corresponds to the impurity collision integral via its action on a function of momentum $\phi_\pp$ :
\begin{equation}\label{eq:Lkernel}
  \hat {\cal L}\circ \phi_\pp=\int (d\pp')w^{\textrm{S}}_{\pp,\pp'}(\phi_{\pp}-\phi_{\pp'})\d(\e_\pp-\e_{\pp'}).
\end{equation}
Using this operator, and working to linear order in the external fields allows to write Eq.~\eqref{eq:intrinsickineq} as
\begin{equation}\label{eq:kineq1}
(i\w-i\qq\p_\pp \e_\pp)g_\pp-\hat{\cal L}\circ g_\pp=e\EE\p_\pp \e_\pp
+\hat {\cal L}\circ\m^{\textrm{int}}_\pp\BB.
\end{equation}
It can be easily shown that $e\EE\p_\pp\e_\pp$ and $\m^{\textrm{int}}_\pp\BB$ do not belong to the kernel of $\hat{\cal L}$, and that $\hat {\cal L}$ is invertible on these functions. Physically, this stems from the fact that both functions are odd in momenta, and the distribution function perturbations that they induce can be relaxed by disorder.

 Then the solution of \eqref{eq:kineq1} is
\begin{eqnarray}
  g^\EE_\pp&=&(i\w-\hat{\cal L})^{-1}\circ e\EE\p_\pp \e_\pp,\nonumber\\
  g^\BB_\pp&=&(i\w-\hat{\cal L})^{-1}\circ\hat {\cal L}\circ\m^{\textrm{int}}_\pp\BB.
\end{eqnarray}

Using $f^{\EE,\BB}_\pp=g^{\EE,\BB}_\pp\p_{\e_\pp}f^0_\pp$, $\p_\pp\e_\pp\p_{\e_\pp}f^0_\pp=\p_\pp f^0_\pp$, as well as Faraday's law $\BB=\qq\times\EE/\w$, the gyrotropic current becomes
\begin{equation}
  \jj_g^{\textrm{int}}(\w,\qq)=e\int(d\pp)\p_\pp f^0_\pp \frac{1}{\w+i \hat{\cal L}}\circ\m^{\textrm{int}}_\pp\cdot\qq\times\EE
  +e\int(d\pp)\qq\times\m^{\textrm{int}}_\pp\frac{1}{\w+i\hat{\cal L}}\circ\EE\p_\pp f^0_\pp.
\end{equation}
The corresponding tensor $\lambda_{abc}$ is given by
\begin{equation}
  \lambda^{\textrm{int}}_{abc}=e\e_{dcb}\int(d\pp)\p_a f^0_\pp \frac{1}{\w+i \hat{\cal L}}\circ m^{\textrm{int}}_{\pp d}
  +e\e_{acd}\int(d\pp)m^{\textrm{int}}_{\pp d}\frac{1}{\w+i\hat{\cal L}}\circ \p_b f^0_\pp.
\end{equation}
This expression can be further modified using the obvious symmetry property of operator $\hat{\cal L}$:
\begin{equation}
  \int(d\pp)\chi_\pp \hat{\cal L}\circ \phi_\pp=\int(d\pp)\phi_\pp \hat{\cal L}\circ \chi_\pp,
\end{equation}
which is inherited by $(\w+i\hat{\cal L})^{-1}$. This brings $\lambda^{\textrm{int}}_{abc}$ to its final form,
\begin{equation}\label{eq:intrinsicfinal}
  \lambda^{\textrm{int}}_{abc}=e\int(d\pp)m^{\textrm{int}}_{\pp d} \frac{1}{\w+i \hat{\cal L}}\circ( \e_{bdc}\p_a\e_\pp -\e_{adc}\p_b\e_\pp )\p_{\e_\pp}f^0_\pp.
\end{equation}
This form makes it apparent that regardless of the explicit form of $w^{\textrm{S}}_{\pp,\pp'}$, the intrinsic contribution to $\lambda_{abc}$ in a disordered metal satisfies $\lambda^{\textrm{int}}_{abc}=-\lambda^{\textrm{int}}_{bac}$, as required by the Onsager relations. In the relaxation-time approximation, $\hat{\cal L}\circ \phi_\pp\to \phi_\pp/\tau$, Eq.~\eqref{eq:intrinsicfinal} reduces to the result of Ref.~\onlinecite{MaPesin2015}.

\subsection{Skew scattering mechanism}
The skew scattering mechanism stems from the asymmetric part of the collision integral, and is determined using the following kinetic equation:
\begin{equation}\label{eq:skewkineq}
-i\w f_\pp+i\qq\p_\pp \e_\pp f_\pp+e\EE\p_\pp f^0_\pp=I^\textrm{S}+I^\textrm{A}.
\end{equation}
The gyrotropic current due to this mechanism is given by the usual ballistic current calculated using the linear-in-$\qq$ part of the solution of this equation, $ f^{\textrm{sk}}_\pp(\qq)$:
\begin{equation}\label{eq:skewcurrent}
   \jj_{\gyr}^{\textrm{sk}}(\w,\qq)=e\int (d\pp)\p_\pp\e_\pp f^{\textrm{sk}}_\pp(\qq).
\end{equation}
Note that even for $I^\textrm{A}=0$ there exists a linear-in-$\qq$ correction to the distribution function as determined by Eq.~\eqref{eq:skewkineq}, but it is straightforward to show that this part of the distribution function does not make a contribution to the current~\eqref{eq:skewcurrent}. Thus $f^{\textrm{sk}}_\pp(\qq)$ is understood as the linear-in-$\qq$ term in the distribution function that exists due to $I^\textrm{A}\neq 0$.

To solve the kinetic equation~\eqref{eq:skewkineq} to linear order in $\qq$, we treat $I^\textrm{A}$ as a perturbation: we solve the $I^\textrm{A}=0$ equation, and substitute the corresponding solution into $I^\textrm{A}$, at which point the latter becomes an effective generation term.  To avoid extremely cumbersome expressions, we use the relaxation time approximation for the operator $\hat{\cal L}$ that defines the symmetric part of the collision integral, introduced in Eq.~\eqref{eq:Lkernel} of Section~\ref{sec:intrinsic}.

In the relaxation time approximation, the solution to Eq.~\eqref{eq:skewkineq} with $I^\textrm{A}=0$ is
\begin{equation}
 f_\pp=\frac{1}{i\left(\w+\frac{i}\tau-\qq\p_\pp \e_\pp\right)}e\EE\p_\pp\e_\pp\p_{\e_\pp} f^0_\pp.
\end{equation}
Substituting this expression back into $I^\textrm{A}$, and using Eq.~\eqref{eq:unitarity} to simplify Eq.~\eqref{eq:skewkineq}, we obtain
\begin{equation}
 f^{\textrm{sk}}_\pp(\qq)=\p_{\e_{\pp}} f^0_{\pp}\int (d\pp') w^{\textrm{A}}_{\pp\pp'}
\frac{\d(\e_{\pp}-\e_{\pp'})}{\left(\w+\frac{i}\tau-\qq\p_\pp \e_\pp\right)\left(\w+\frac{i}\tau-\qq\p_{\pp'} \e_{\pp'}\right)}e\EE\p_{\pp'}\e_{\pp'},
\end{equation}
where we used the fact that $\d(\e_{\pp}-\e_{\pp'})\p_{\e_{\pp'}} f^0_{\pp'}=\d(\e_{\pp}-\e_{\pp'})\p_{\e_{\pp}} f^0_{\pp}$. Separating the linear-in-$\qq$ part in $f^{\textrm{sk}}_\pp(\qq)$, and substituting it into Eq.~\eqref{eq:skewcurrent} for the current, we obtain
\begin{equation}
   \jj_{\gyr}^{\textrm{sk}}(\w,\qq)=\frac{e^2}{(\w+\frac{i}\tau)^3}\int (d\pp)\int (d\pp')w^{\textrm{A}}_{\pp\pp'}
\d(\e_{\pp}-\e_{\pp'})\left(\qq\p_\pp \e_\pp+\qq\p_{\pp'} \e_{\pp'}\right)(\EE\p_{\pp'}\e_{\pp'})\p_\pp f^0_{\pp}.
\end{equation}
The corresponding contribution to tensor $\lambda_{abc}$ is
\begin{equation}
   \lambda^{\textrm{sk}}_{abc}=\frac{e^2}{(\w+\frac{i}\tau)^3}\int (d\pp)\int (d\pp')w^{\textrm{A}}_{\pp\pp'}
\d(\e_{\pp}-\e_{\pp'})\p_{a}\e_{\pp}\p_{b}\e_{\pp'} \left(\p_c \e_\pp+\p_{c} \e_{\pp'}\right)\p_{\e_\pp} f^0_{\pp}.
\end{equation}
The antisymmetry of $w^{\textrm{A}}_{\pp\pp'}$ ensures that this tensor is antisymmetric with respect to the first pair of indices.

The result for $\lambda^{\textrm{sk}}_{abc}$ can be written more compactly if one introduces the ``skew acceleration'' $\bm Q_\pp$, which has the meaning of the rate of change of an electron's velocity due to the skew scattering event accumulation:
\begin{equation}\label{eq:skacceleration}
  \bm Q^{\textrm{sk}}_\pp=\int (d\pp')w^{\textrm{A}}_{\pp\pp'}
\d(\e_{\pp}-\e_{\pp'})(\p_{\pp}\e_{\pp}-\p_{\pp'}\e_{\pp'}).
\end{equation}
Note that the term with $\p_\pp\e_\pp$ in the round brackets integrates to zero due to the probability flux conservation condition~\eqref{eq:unitarity}, and has been added for clarity of the physical interpretation.
This allows to write finally
\begin{equation}\label{eq:skewfinal}
   \lambda^{\textrm{sk}}_{abc}=\frac{e^2}{(\w+\frac{i}\tau)^3}\int (d\pp)
   \left(\p_{a}\e_{\pp}Q^{\textrm{sk}}_{\pp b}-Q^{\textrm{sk}}_{\pp a}\p_{b}\e_{\pp}\right)\p_c \e_\pp\p_{\e_\pp} f^0_{\pp}.
\end{equation}

\subsection{Side jump mechanism}
The side jump contribution to the gyrotropic current comes from the following kinetic equation:
%\begin{eqnarray}
% -i\w f_\pp+i\qq\p_\pp \e_\pp f_\pp+e\EE\p_\pp f^0_\pp&=&-\int (d\pp') W^{s}_{\pp,\pp'}(f_{\pp}-f_{\pp'})\d(\e_\pp-\e_{\pp'})-\p_{\e_\pp} f^0_\pp\int (d\pp')W^s_{\pp\pp'}e\EE\d\rr_{\pp\pp'}\delta\left(\e_\pp-\e_{\pp'}\right),
%\end{eqnarray}
\begin{eqnarray}
 -i\w f_\pp+i\qq\p_\pp \e_\pp f_\pp+e\EE\p_\pp f^0_\pp&=&I^{\textrm{S}}+I^{\textrm{sj}},
\end{eqnarray}
supplemented with the expression for the current that contains the contribution from the side jump accumulation:
\begin{equation}\label{eq:gyrosj}
  \jj_{\gyr}^{\textrm{sj}}(\w,\qq)=e\int (d\pp)\p_\pp\e_\pp f^{\textrm{sj}}_\pp(\qq)+e\int(d\pp)\int(d\pp')w^{\textrm{S}}_{\pp\pp'}\d\rr_{\pp\pp'} f^{\EE}_{\pp'}(\qq)\delta\left(\e_{\pp'}-\e_{\pp}\right).
\end{equation}
Here $f^{\textrm{sj}}_\pp(\qq)$ is the linear-in-$\qq$ correction to the distribution function due to the presence of side jumps in the external electric field, while $ f^{\EE}_\pp(\qq)$ is the linear-in-$\qq$ correction to the distribution function in the absence of side jumps.

As before, we introduce $f^{\EE}_{\pp}(\qq)=g^{\EE}_{\pp}(\qq)\p_{\e_\pp}f^0_\pp$, and  $f^{\textrm{sj}}_\pp(\qq)=g^{\textrm{sj}}_\pp(\qq)\p_{\e_\pp}f^0_\pp$, and treat the symmetric part of the collision integral in the relaxation time approximation. Then for the $g$'s we obtain in the standard way
\begin{equation}
  g^{\EE}_\pp(\qq)=\frac{1}{i(\w+\frac{i}{\tau})^2}(\qq\p_{\pp}\e_\pp)(e\EE\p_{\pp}\e_\pp),
\end{equation}
and
\begin{equation}
  g^{\textrm{sj}}_\pp(\qq)=\frac{1}{i(\w+\frac{i}{\tau})^2}(\qq\p_{\pp}\e_\pp)\int (d\pp')w^{\textrm{S}}_{\pp\pp'}e\EE\d\rr_{\pp\pp'}\delta\left(\e_\pp-\e_{\pp'}\right).
\end{equation}
Now we can calculate the gyrotropic part of the current due to the side jump mechanism, Eq.~\eqref{eq:gyrosj}.
\begin{eqnarray}
  \jj_{\gyr}^{\textrm{sj}}(\w,\qq)&=&\frac{e}{i(\w+\frac{i}{\tau})^2}\int (d\pp)\int (d\pp')w^{\textrm{S}}_{\pp\pp'}\p_\pp\e_\pp(\qq\p_{\pp}\e_\pp)(e\EE\d\rr_{\pp\pp'})\delta\left(\e_\pp-\e_{\pp'}\right)\p_{\e_\pp}f^0_\pp\nonumber\\
  %%%%%%%%%%%%%%%
  &+&\frac{e}{i(\w+\frac{i}{\tau})^2}\int(d\pp)\int(d\pp')w^{\textrm{S}}_{\pp\pp'}\d\rr_{\pp\pp'} (\qq\p_{\pp'}\e_{\pp'})(e\EE\p_{\pp'}\e_{\pp'})\delta\left(\e_{\pp}-\e_{\pp'}\right)\p_{\e_{\pp}}f^0_{\pp}.
\end{eqnarray}
By interchanging $\pp$ and $\pp'$ in the second term, and using $\d\rr_{\pp'\pp}=-\d\rr_{\pp\pp'}$, the above expression for the gyrotropic current can be seen to be equivalent to the following contribution to tensor $\lambda_{abc}$:
\begin{eqnarray}
  \lambda^{\textrm{sj}}_{abc}&=&\frac{e^2}{i(\w+\frac{i}{\tau})^2}\int (d\pp)\int (d\pp')w^{\textrm{S}}_{\pp\pp'} \delta\left(\e_\pp-\e_{\pp'}\right)(\p_a\e_\pp \d \rr_{\pp\pp',b}-\d\rr_{\pp\pp',a} \p_{b}\e_{\pp}) \p_c f^0_\pp,
\end{eqnarray}
which is obviously antisymmetric with respect to the first pair of indices.

The expression for $\lambda^{\textrm{sj}}_{abc}$ can be written in a more compact form if we introduce the side jump accumulation velocity according to
\begin{equation}\label{eq:sjvelocity}
  \vv^{\textrm{sja}}_\pp=\int (d\pp')w^{\textrm{S}}_{\pp\pp'}\d \rr_{\pp\pp'} \delta\left(\e_\pp-\e_{\pp'}\right).
\end{equation}
This brings $\lambda^{\textrm{sj}}_{abc}$ to its final form:
\begin{eqnarray}\label{eq:sjfinal}
  \lambda^{\textrm{sj}}_{abc}&=&\frac{e^2}{i(\w+\frac{i}{\tau})^2}\int (d\pp)(\p_a\e_\pp \vv^{\textrm{sja}}_{\pp,b}-\vv^{\textrm{sja}}_{\pp,a} \p_{b}\e_{\pp}) \p_c\e_\pp \p_{\e_\pp} f^0_\pp.
\end{eqnarray}
\end{widetext}

Eqs.~\eqref{eq:intrinsicfinal},~\eqref{eq:skewfinal}, and~\eqref{eq:sjfinal} for intrinsic, skew scattering, and side jump contributions to tensor $\lambda_{abc}$ are one of the main results of this paper. The physical origin and the meaning of the extrinsic mechanisms are elaborated upon in the next Section.

\section{Physical consequences of the obtained results}\label{sec:physics}

In this section we discuss the physical content of the tensor $\lambda_{abc}$. To this end, we focus on the kinetic magnetoelectric effect, and natural optical activity. The main goal is to elucidate the relative role of the intrinsic and extrinsic mechanisms of nonlocality. We will show that skew scattering dominates in clean noncentrosymmetric metals -- a familiar observation from the theory of the AHE. However, there are realistic examples of materials (e.g. Tellurium and TaAs-family Weyl semimetals) in which intrinsic effects dominate in experimentally available samples.

For continuity of the presentation, we defer the question of symmetry requirements for the physical phenomena we discuss till the end of this Section (see Section~\ref{sec:symmetry}).

\subsection{Gyrotropic tensor}\label{sec:gyrotensor}
Being antisymmetric with respect to the first pair of indices, $\lambda_{abc}$ contains nine independent components, and hence is dual to a second rank pseudotensor:
\begin{equation}\label{eq:lambdatog}
  \lambda_{abc}=\e_{abd}g_{dc}.
\end{equation}
In what follows, we call $g$, defined by Eq.~\eqref{eq:lambdatog}, the gyrotropic tensor. Explicitly, it is given by
\begin{equation}\label{eq:gtensor}
  g_{ab}=\frac12\e_{cda}\lambda_{cdb}.
\end{equation}
It is customary and simpler to use the gyrotropic tensor to describe the physical properties of a crystal. The intrinsic, skew scattering, and side jump contributions to $g_{ab}=g_{ab}^{\textrm{int}}+g_{ab}^{\textrm{sk}}+g_{ab}^{\textrm{sj}}$ can be read off from Eqs.~\eqref{eq:intrinsicfinal},~\eqref{eq:skewfinal} and~\eqref{eq:sjfinal}. Using  $\p_{a}\e_\pp\p_{\e_\pp}f^0_\pp=\p_{a}f^0_\pp$ and Eq.~\eqref{eq:gtensor}, we obtain
\begin{subequations}
\label{eq:gtensorall}
\begin{align}
  g_{ab}^{\textrm{int}}&= \frac{e}{(\w+\frac{i}{\tau})}\int(d\pp)
  (m_{\pp a} \p_bf^0_\pp -\d_{ab}\m_{\pp}\cdot\p_\pp f^0_\pp),\label{eq:intg}\\
  %%%%%%%%%%%%%%%%%%%%%%%%%%
  %%%%%%%%%%%%%%%%%%%%%%
  %%%%%%%%%%%%%%%%%%
  g_{ab}^{\textrm{sk}}&= -\frac{e^2}{(\w+\frac{i}\tau)^3}\int (d\pp)
[\bm Q^{\textrm{sk}}_\pp\times\p_{\pp}\e_{\pp}]_a\p_b f^0_{\pp},\label{eq:skewg}\\
%%%%%%%%%%%%%%%%%%%%%%%%%%%%
%%%%%%%%%%%%%%%%%%%%%%%
%%%%%%%%%%%%%%%%%%
  g_{ab}^{\textrm{sj}}&= -\frac{e^2}{i(\w+\frac{i}{\tau})^2}\int (d\pp)[\vv^{\textrm{sja}}_{\pp} \times\p_\pp\e_\pp]_a \p_b f^0_\pp.\label{eq:sjg}
\end{align}
\end{subequations}
The expressions for the intrinsic orbital moment of quasiparticles, $\m_{\pp}$, the skew acceleration, $\bm Q^\textrm{sk}_\pp$, and the side jump accumulation velocity $\vv^{\textrm{sja}}_{\pp}$ are given by Eqs.~\eqref{eq:magneticmoment},~\eqref{eq:skacceleration}, and~\eqref{eq:sjvelocity}, respectively.

Eqs.~\eqref{eq:gtensorall} make it obvious that both intrinsic and extrinsic contributions to the gyrotropic tensor have very similar structure. This similarity can be further emphasized by introducing the effective kinetic orbital magnetic moment $\m^{\textrm{kin}}_\pp$:
\begin{equation}\label{eq:kinmoment}
  \m^{\textrm{kin}}_\pp=\frac{e\tau^2}{(1-i\w\tau)^2}\bm Q^{\textrm{sk}}_\pp\times\p_{\pp}\e_{\pp}
  +\frac{e\tau}{1-i\w\tau}\vv^{\textrm{sja}}_{\pp} \times\p_\pp\e_\pp,
\end{equation}
the two terms in which correspond to the skew scattering and side jump contributions to the orbital moment, see Eqs.~\eqref{eq:skmomentresults} and~\eqref{eq:sjmomentresults} in Section~\ref{sec:results} and the text around them for further discussion. Using the total magnetic moment $\m^{\textrm{tot}}_\pp=\m^{\textrm{int}}_\pp+\m^{\textrm{kin}}_\pp$, and noting that  $\bm Q^{\textrm{sk}}_\pp\times\p_{\pp}\e_{\pp}\cdot\p_\pp f^0_{\pp}=0$ and $\vv^{\textrm{sja}}_{\pp} \times\p_\pp\e_\pp\cdot \p_\pp f^0_\pp=0$, the entire gyrotropic tensor -- the sum of Eqs.~\eqref{eq:gtensorall}--can be written as
\begin{equation}\label{eq:gyrototal}
  g_{ab}= \frac{e}{(\w+\frac{i}{\tau})}\int(d\pp)
  (m^{\textrm{tot}}_{\pp a} \p_bf^0_\pp -\d_{ab}\m^{\textrm{tot}}_{\pp}\cdot\p_\pp f^0_\pp).
\end{equation}
in complete analogy with the intrinsic contribution, Eq.~\eqref{eq:intg}.

We note in passing that the kinetic part of the magnetic moment, Eq.~\eqref{eq:kinmoment}, does not make a contribution to the trace of the gyrotropic tensor,
\begin{equation}
  \Tr g=-\frac{2e}{(\w+\frac{i}{\tau})}\int(d\pp)
  \m_{\pp}\cdot\p_\pp f^0_\pp.
\end{equation}
This can be shown to be a consequence of our restricting treatment of the intrinsic effects to the linear order in skew scattering probability, and the side jump length. This point is further elaborated upon at the end of Section~\ref{sec:symmetry}.

\subsection{Natural optical activity and Faraday rotation}\label{sec:NOA}

One of physical consequences of the linear-in-$\qq$ spatial dispersion of the conductivity tensor is the phenomenon of natural optical activity, whereby a crystal responds differently to right- and left- circularly polarized light. The difference in real parts of the refractive indices results in polarization rotation upon transmission through a sample, and the difference in the imaginary parts, and hence absorption coefficients, leads to the circular dichroism. The goal of the present Section is to consider the contribution of extrinsic effects to the natural optical activity in metals. The intrinsic ones have been previously studied in Refs.~\onlinecite{MaPesin2015,Zhong2016}.

Propagation of electromagnetic waves through a crystal is governed by the equivalent dielectric tensor $\ve_{ab}(\w,\qq)$ that corresponds to the conductivity tensor~\eqref{eq:conductivity_general}:
\begin{align}
  \ve_{ab}(\w,\qq)=\d_{ab}+\frac{i}{\w\ve_0}\sigma_{ab}(\w,\qq).\nonumber
\end{align}
In general, in an anisotropic crystal the combined  effect of optical activity and birefringence leads to tedious considerations of wave propagation. All the details can be found in textbooks\cite{Malgrange2014}. Here, we restrict ourselves to the simple case of  propagation along the optic axis of a uniaxial crystal with point group $D_3$. We did not choose a more symmetric group (like $O$), in which the gyrotropic tensor would be proportional to the unit tensor, since the extrinsic part of the gyrotropic tensor in Eq.~\eqref{eq:gyrototal} is traceless, and hence vanishes in isotropic crystals.

In a crystal with $D_3$ point group, the gyrotropic and local conductivity tensors are diagonal:
\begin{align}
  [g,\sigma]=\begin{pmatrix}
    [g,\sigma]_{xx}&0&0\\
    0&[g,\sigma]_{xx}&0\\
    0&0&[g,\sigma]_{zz}
  \end{pmatrix}
\end{align}

For propagation along the optic axis ($z$-axis), the wave equation reads
\begin{align}\label{eq:maxwell}
  \left[q^2\left(\d_{ab}-\frac{q_aq_b}{q^2}\right)-\frac{\w^2}{c^2}\ve_{ab}(\w,\qq)\right]E_b=0,
\end{align}
 and the electric field of the wave has only x- and y-components.  The eigenvectors of the $2\times2$ matrix equation \eqref{eq:maxwell} are $(1,\pm i)^T$ ($T$ - transposition), and correspond to the left and right circular polarizations of light. For a given $\w$, the solutions for the corresponding wave numbers are given to linear order in $g_{zz}$ by
\begin{align}
q_{L,R}=\frac{\w}{c}\sqrt{1+\frac{i\sigma_{xx}(\w)}{\w\ve_0}}\mp \frac{1}{2}\mu_0\w g_{zz} .\nonumber
\end{align}

Half of the difference between these wave numbers defines the complex rotatory power of the crystal, observed in Faraday rotation for transmitted light:
\begin{align}
  \rho=\frac{1}{2}\mu_0\w g_{zz}.
\end{align}

To estimate the rotatory power due to the extrinsic effects, we make several assumptions. First, we assume that the skew scattering mechanism dominates over the side jump one, as is common under realistic circumstances. Second, we consider a metal with Fermi energy $\e_F$, Fermi velocity $v_F$, Fermi momentum $p_F$, effective mass $m$, and the density of states at the Fermi level $\nu_F\sim m p_F/\hbar^3$. Note that the following estimates also work for Weyl metals, in which case one should set $m\sim p_F/v_F$. Third, we introduce the skew scattering time $\tau_{\textrm{sk}}$ via the typical value of the skew acceleration, Eq.~\eqref{eq:skacceleration}, as $Q^{\textrm{sk}}\sim v_F/\tau_{\textrm{sk}}$. Finally, in order to avoid the necessity to discriminate between the reactive and dissipative effects, we consider $\w\sim 1/\tau$, where these effects are of the same order. Then Eq.~\eqref{eq:skewg} yields
\begin{align}
  g_{zz}\sim  \frac{e^2}{(\w+\frac{i}\tau)^3}\frac{\nu_F v_F^3}{\tau_{\textrm{sk}}},
\end{align}
and the corresponding contribution to the rotatory power at $\w\sim 1/\tau$ becomes
\begin{align}
  \rho\sim \frac{1}{c\tau}\frac{e^2}{\ve_0\hbar c}\frac{\tau}{\tau_{\textrm{sk}}}(k_F\ell)^2,
\end{align}
where $c$ is the speed of light. For $k_F\ell\sim 10$, $\tau/\tau_{\textrm{sk}}\sim 0.01$, and $\tau\sim 1$ps, we obtain $\rho\sim 0.1$rad/mm. This is a considerable number for THz-range rotatory power, comparable to the rotatory power of alpha-quartz at optical frequencies, or clean Weyl metal in the infrared\cite{Zhong2016}.

\subsection{Kinetic magnetoelectric effect and the current-induced magnetization}\label{sec:kineticME}

Magnetoelectric phenomena are usually discussed in the context of materials with broken time-reversal symmetry, in which electric fields cause magnetic response, and vice versa. However, even in time-reversal invariant  (yet noncentrosymmetric) crystals the response of the magnetization to an electric field is possible. In that case, different time-reversal parity of quantities in the left and right hand sides of a linear relationship $M\propto E$ implies that there must be a dissipative process underlying the response, that is, the current flow. Therefore, such kinetic magnetoelectric response~\cite{Levitov,Zhong2016} in time-reversal systems can also be viewed as the phenomenon of current-induced magnetization. Below we will show that the dc limit of such a response is completely determined by the tensor $\lambda_{abc}$, or equivalently, the gyrotropic tensor.

Physically, such a conclusion is based on the expression \eqref{eq:gyrototal} for the total gyrotropic tensor: its similarity to the expression for the intrinsic contribution, Eq.~\eqref{eq:intg}, led us to identify the right hand side of Eq.~\eqref{eq:kinmoment} as an effective kinetic magnetic moment of quasiparticles. If this identification is correct, one expects that the current-induced magnetization will be related to a finite density of the total magnetic moment, given by the sum of the intrinsic, Eq.~\eqref{eq:magneticmoment}, and extrinsic, Eq.~\eqref{eq:kinmoment}, contributions. Below, by arriving at this conclusion from the macroscopic electrodynamics point of view, we will confirm the  interpretation of Eq.~\eqref{eq:kinmoment} once again.

To determine the magnetization response to a dc transport current, we consider the ``$q\to 0$ first, $\w\to 0$ second'' limit of linear response to electromagnetic fields. In this limit, the response of a crystal is fully determined by the macroscopic  electric polarization and magnetization\cite{BarronBook,Melrose,Malashevich}, while the quadrupolar and higher order polarizations, of both electric and magnetic character, can be neglected for the purpose of considering linear-in-$\qq$ spatial dispersion effects.

The general response of polarization and magnetization to electromagnetic fields is written as
\begin{subequations}\label{eq:magnetoelectricresponse}
  \begin{align}
    P_a(\w,\qq)&=\chi^\textrm{e}_{ab}(\w)E_b(\w,\qq)+i\chi^\textrm{em}_{ab}(\w)B_b(\w,\qq),\\
    M_a(\w,\qq)&=-i\chi^\textrm{me}_{ab}(\w)E_b(\w,\qq)+\chi^\textrm{m}_{ab}(\w)B_b(\w,\qq).
  \end{align}
\end{subequations}
We neglected the $\qq$ dependence of the response tensors $\chi^{\textrm{e},\textrm{m},\textrm{em},\textrm{me}}$: this approximation is sufficient to discuss the effects of spatial dispersion to linear order in $\qq$. The magnetoelectric susceptibility $\chi^{\textrm{me}}_{ab}$ describes the magnetization response to a transport electric field, the so-called kinetic magnetoelectric effect\cite{Levitov}. Our theory allows to compare the intrinsic and extrinsic contributions to such current-induced magnetization.

To relate $\chi^{\textrm{me}}_{ab}$ to the gyrotropic tensor $g_{ab}$, we note that the gyrotropic current is the linear-in-$\qq$ part of the macroscopic electric current, $\jj=-i\w \bm P(\w,\qq)+i\qq\times \bm M(\w,\qq)$, given by
\begin{equation}\label{eq:macrocurrent}
  j_{\gyr,a}=(\chi^{\textrm{em}}_{ad}\epsilon_{dcb}+\e_{acd} \chi^{\textrm{me}}_{db})q_c E_b.
\end{equation}
Combined with the Onsager symmetry relation for the magnetoelectric response tensors,
\begin{equation}
    \chi^\textrm{em}_{ab}(\w)=\chi^\textrm{me}_{ba}(\w),
  \end{equation}
Eq.~\eqref{eq:macrocurrent} gives the following expression for $\lambda_{abc}$ in Eq.~\eqref{eq:conductivity_general}:
\begin{equation}
  \lambda_{abc}=(\e_{acd} \chi^{\textrm{me}}_{db}-\epsilon_{bcd}\chi^{\textrm{me}}_{da}).
\end{equation}
This allows to express the gyrotropic tensor via the mangetoelectric one:
\begin{eqnarray}
  g_{ab}&=&\frac{1}{2}\e_{cda}\lambda_{cdb}=\chi^{\textrm{me}}_{ab}-\d_{ab}\Tr\chi^{\textrm{me}}.
\end{eqnarray}
Conversely, using that $\Tr g=-2\Tr \chi^{\textrm{me}}$, one obtains
\begin{equation}\label{eq:magnetoelectrictensor}
  \chi^{\textrm{me}}_{ab}=g_{ab}-\frac{1}{2}\d_{ab}\Tr g.
\end{equation}
This expression solves the problem of expressing the magnetoelectric susceptibility through the microscopically calculated gyrotropic tensor. We note that for an isotropic system $g_{ab}=g\d_{ab}$, and Eq.~\eqref{eq:magnetoelectrictensor} yields  $\chi^{\textrm{me}}_{ab}=-\frac{1}{2}g\d_{ab}$, in agreement with previously known results\cite{Levitov}.

One can make another step to bring out the essence of the phenomenon. To this end, we use Eq.~\eqref{eq:gyrototal} for the gyrotropic tensor to obtain
\begin{align}
  \chi^{\textrm{me}}_{ab}=\frac{e}{(\w+\frac{i}{\tau})}\int(d\pp)
  m^{\textrm{tot}}_{\pp a} \p_bf^0_\pp ,\nonumber
\end{align}
which gives the following expression for the magnetization:
\begin{align}
  \bm M=\int(d\pp)
 (\m^{\textrm{int}}_\pp+\m^{\textrm{kin}}_\pp)\frac{e\EE\cdot\p_\pp f^0_\pp}{(i\w-\frac{1}{\tau})}.
\end{align}
The meaning of this expression is obvious: the second factor inside the momentum space integral is the change in the electronic distribution function under the action of an electric field, and the term in the brackets is the effective magnetic moment of quasiparticles, which includes both the intrinsic part, $\m^{\textrm{int}}_\pp$ of Eq.~\eqref{eq:magneticmoment}, and extrinsic - kinetic - contribution, $\m^{\textrm{kin}}_\pp$ of Eq.~\eqref{eq:kinmoment}. This again confirms our interpretation of Eq.~\eqref{eq:kinmoment}.

A detailed study of the current-induced magnetization in three-dimensional metals will be presented in a separate work\cite{RouUnpublished}, but it is obvious from Eq.~\eqref{eq:kinmoment} that in a clean enough sample the skew scattering contribution to the kinetic momentum will always dominate the effect, since it is inversely proportional to the impurity concentration, unlike the intrinsic and side-jump contributions, which are independent of the impurity concentration.

The phenomenon of current-induced magnetization has mostly been considered in two-dimensional systems\cite{Edelstein1990,Kato2004magnetization,Sih2005,Stern2006current}, where it is of spin origin, simply because reduced dimensionality does not allow electrons to ``orbit'' around the direction of the current flow. For three-dimensional helical metals, there are theoretical\cite{Murakami2015}  studies of the intrinsic mechanism (both of orbital and spin character) of the current-induced magnetization. On the experimental side, the intrinsic contribution to current-induced magnetization was considered in Ref.~\onlinecite{Farbshtein2012} in Tellurium. Here, in order to have a rough estimate of the relative magnitudes of extrinsic and intrinsic contributions, we present a comparison of intrinsic and skew scattering mechanisms of current-induced magnetization in relatively low mobility ($\textrm{3300 V/cm s}^2$) Tellurium samples used in the experiments of Ref.~\onlinecite{Farbshtein2012}.

Near the top of the valence band of p-doped Tellurium, the intrinsic magnetic moments of quasiparticles point along the optical axis of the crystal, and can be estimated as\cite{Farbshtein2012}
\begin{equation}
  m^{\textrm{int}}\sim \frac{e\hbar}{m^*}\frac{\beta k_T}{\D},
\end{equation}
where $m^*\sim 0.1 m_0$, $m_0$ being the bare electron's mass, $\beta=2.4\times 10^{-10}$ eV$\cdot$m is a velocity related to the spin-orbit coupling, $\Delta=126$meV is the gap between the two top valence bands, and $k_T\sim 10^{8}\textrm{m}^{-1}$ is the thermal wave number at $T=77$K.

In turn, the skew-scattering part of the kinetic magnetic moment can be estimated by setting the skew-scattering acceleration to be $Q^{\textrm{sk}}\sim v_T/\tau_{\textrm{sk}}$, where $\tau_{\textrm{sk}}$ is the skew scattering time, and $v_T$ is the thermal velocity of carriers. This yields
\begin{equation}
  m^{\textrm{kin}}\sim \frac{e\tau^2}{\tau_{\textrm{sk}}}v_T^2.
\end{equation}
Taking the ratio of the last two equations, we obtain
\begin{equation}
  \frac{m^{\textrm{kin}}}{m^{\textrm{int}}}\sim \frac{\D \tau}{\hbar}\frac{m^* v_T^2}{\beta k_T}\frac{\tau}{\tau_{\textrm{sk}}}.
\end{equation}
Since the ratio $\tau/\tau_{\textrm{sk}}$ is independent of the impurity concentration, and $\tau$ is inversely proportional to it, the ratio becomes large for clean samples. This, of course, is a manifestation of the general trend that extrinsic effects dominate in clean materials\cite{NagaosaReview}. For samples used in Ref.~\onlinecite{Farbshtein2012}, $\tau\sim 6\times 10^{-13}$s was reported, and one can also estimate\cite{RouUnpublished} $\tau/\tau_{\textrm{sk}}\sim 0.001-0.01$, which makes $m^{\textrm{kin}}/m^{\textrm{int}}\sim 0.1 - 1$. That is, the role of the kinetic moments is substantial even relatively dirty samples.

Note also that for electric fields perpendicular to the optical axis of a Te crystal, the intrinsic contribution vanishes identically, yet the extrinsic ones do not. Hence for such field directions the extrinsic effects completely determine the current-induced magnetization\cite{RouUnpublished}.

We can also present an estimate of the extrinsic contribution to the current-induced magnetization in a clean, $k_F\ell\sim100$, and ``strongly helical'', $\tau/\tau_{\textrm{sk}}\sim 0.01$, metal. Using the estimate for the skew contribution to the gyrotropic tensor, Eq.~\eqref{eq:skgyroresults}, and thetransport electric field of $10^4$V/m, we obtain $\mu_0 M\sim 5$ Gauss. If this metal were a typical noncentrosymmetric Weyl metal with $v_F\sim 10^6$m/s, then an extrinsic contribution of 5 Gauss per $10^4$V/m of electric field would be equivalent to the Weyl point separation in energy of 0.1eV - a figure typical for all known examples of time-reversal invariant Weyl metals.

\subsection{Symmetry requirements}\label{sec:symmetry}
To complete the discussion of physical effects related to the linear-in-$\qq$ spatial dispersion of the conductivity tensor, below we list the point groups of metals in which one expects the phenomena of natural optical activity and current-induced magnetization to happen. Both require broken inversion symmetry, which is necessary to have a nonzero third rank tensor ($\lambda_{abc}$). As far as the current-induced magnetization is concerned, having nonzero ($\lambda_{abc}$), or equivalently, $g_{ab}$ tensors is in fact the only requirement on the crystal symmetry, as follows from Eq.~\eqref{eq:magnetoelectrictensor}. Therefore, of the 21 noncentrosymmetric point groups (which can be found in Ref.~\onlinecite{Nye}, and will not be listed here), 18 are gyrotropic, and allow at least current-induced magnetization, but not necessarily natural optical activity. The three noncentrosymmetric groups that have zero $\lambda_{abc}$ are $C_{3h}$, $D_{3h}$, and $T_d$.

It is known\cite{Malgrange2014} that only the symmetric part of the gyrotropic tensor contributes to the natural optical activity. Therefore, besides the nonsymmetricity, one has another condition that the gyrotropic tensor $g_{ab}$ has nonzero symmetric part. Among the 18 gyrotropic groups only 15 satisfy this condition, with $C_{3v}$, $C_{4v}$, and  $C_{6v}$ being the exceptions.

In order to have robust dynamic Chiral Magnetic Effect, which survives in polycrystalline samples\cite{MaPesin2017}, the gyrotropic tensor of a crystal must have a nonzero trace. This further reduces the list of possible groups to 11 chiral ones, which describe enantiomorphic crystals. Such crystals do not possess any mirror symmetries. The four groups that do allow natural optical activity, but do not have robust dynamic Chiral Magnetic Effect are $C_{2h}$, $C_{2v}$, $S_4$, and $D_{2d}$. $S_4$ is excluded as having a roto-reflection axis, rather than a mirror symmetry.

Finally, we would like to mention that the extrinsic contributions to the gyrotropic tensor calculated in this paper vanish in cubic crystals with point groups $T$ and $O$. Indeed, in these groups $g_{ab}$ is proportional to the unit tensor, and since the traces of the extrinsic contributions, Eqs.~\eqref{eq:skewg} and~\eqref{eq:sjg}, vanish, they in fact vanish identically. This can be traced back to our perturbative treatment of the skew scattering and side jump processes, and does not depend on the relaxation time approximation used to simplify the results.

By going to the next order in $w^\textrm{A}$ and $\d \rr_{\pp\pp'}$, one can obtain nonzero extrinsic contributions. This was done for an isotropic single-band crystal with chiral impurities in Ref.~\onlinecite{Levitov}. In that case, the result for the skew scattering contribution, the larger of the two extrinsic ones, contains an additional small factor of $\tau/\tau_\textrm{sk}$. In realistic enantiomorphic cubic crystals, that contribution is likely to be dominated by the intrinsic one\cite{MaPesin2015}. By the same isotropy token, the intrinsic effects will also dominate in polycrystalline samples.

\section{Conclusions}\label{sec:conclusions}
We have considered the leading nonlocal corrections to the conductivity tensor in noncentrosymmetric disordered metals, which are linear in the wave vector of an electromagnetic perturbation. Our discussion pertains to general multiband systems, but is limited by the frequencies which are small compared to all band splittings at the Fermi surface. The relation between the frequency and the elastic scattering rate is arbitrary.

The main result of this work is the identification of two disorder-induced contributions to the magnetic moment of charge carriers, which come from the skew scattering, and side jump effects. These contributions appear only in the nonequilibrium response to the electromagnetic fields (``$\qq\to 0$ first'' limit of response), hence we called them kinetic magnetic moments. Relevant expressions are summarized in Section~\ref{sec:results}.

Physically, the conductivity tensor calculated here describes the phenomena of natural optical activity and the dynamic chiral magnetic effect, as well as the kinetic magnetoelectric effect and the current-induced magnetization. We estimated that in the THz frequency range, a typical helical metal should have rotatory power of $0.1$ rad/mm due to the skew scattering part of the magnetic moment. At low frequencies, $\w\tau\ll1$, the current-induced magnetization in such a metal is around 5 Gauss per $10^4$V/m of applied electric field. These figures show that the extrinsic effects -- mostly related to skew scattering -- are comparable to the intrinsic ones in realistic samples, and both types of effects are expected to be straightforward to measure.

\acknowledgments
This work was supported by the National Science Foundation Grant No. DMR-1409089.

\bibliography{extrinsic_references}

\begin{thebibliography}{61}
\bibitem[{not({\natexlab{a}})}]{noteonspin}
\bibinfo{note}{It can be shown~\cite{Zhong2016} that spin effects can be taken
  into account by simply changing $\m_{n\pp}\to \m_{n\pp}+\bra u_{n\pp}|\bm
  s|u_{n\pp}\ket$, where $\bm s =\hbar\bm \sigma/2$ is the operator of the
  electron's spin}.
\bibitem[{not({\natexlab{b}})}]{noteonunitarity}
\bibinfo{note}{It may seem that taking into account electromagnetic fields in
  the energy-conserving $\d$-function warrants a revision of the
  condition~\eqref{eq:unitarity} imposed by unitarity of scattering. However,
  such effects represent an interference between the skew scattering, and
  semiclassical effect desribed by the wave packet intrinsic orbital magnetic
  moment and the side jump. Such extremely small corrections are neglected in
  this paper.}
\end{thebibliography}
\bibliographystyle{apsrev}

\end{document}